\documentclass[aps,prl,twocolumn]{revtex4-1}
\usepackage[usenames]{color}
\DeclareFontFamily{OT1}{pzc}{}
\DeclareFontShape{OT1}{pzc}{m}{it}{<-> s * [1.10] pzcmi7t}{}
\DeclareMathAlphabet{\mathpzc}{OT1}{pzc}{m}{it}
\newcommand{\comment}[1]{}
\usepackage{graphicx}
\usepackage{epsfig}
\usepackage{amsmath}
\usepackage{mathrsfs}
\usepackage{bm}
\usepackage{color}

\newcommand{\Hc}{{\cal H}}
\newcommand{\Uc}{{\cal U}}
\newcommand{\bk}{{\bf k}}
\newcommand{\bb}{{\bf b}}
\newcommand{\bn}{{\bf n}}
\newcommand{\bu}{{\bf u}}
\newcommand{\bv}{{\bf v}}
\newcommand{\xh}{{\bf \hat{x}}}
\newcommand{\yh}{\bf \hat{y}}
\newcommand{\kx}{{k_x}}
\newcommand{\ky}{{k_y}}
\newcommand{\bsigma}{{\bm \sigma}}
\newcommand{\blambda}{{\bm \lambda}}
\newcommand{\nc}{\mathpzc{n}}

\begin{document}
\title{SU$(3)$ Spin-Orbit Coupling in Systems of Ultracold Atoms}
\date{\today}
\author{Ryan Barnett,$^{1,2,3}$  G. R. Boyd,$^2$ and Victor Galitski$^{1,2}$}
\affiliation{Joint Quantum
Institute$^1$ and Condensed Matter Theory Center$^2$, 
Department of Physics, University of Maryland, College
Park, Maryland 20742-4111, USA}
\affiliation{$^3$Department of Mathematics, Imperial College London,
London SW7 2BZ, United Kingdom}

\begin{abstract}
Motivated by the recent experimental success in realizing synthetic
spin-orbit coupling in ultracold atomic systems, we consider
$N$-component atoms coupled to a non-Abelian SU$(N)$ gauge field. More
specifically, we focus on the case, referred to  here as ``SU(3)
spin-orbit-coupling,''  where the internal states of three-component
atoms are coupled to their momenta via a matrix structure that
involves the Gell-Mann matrices (in contrast to the Pauli matrices in
conventional SU(2) spin-orbit-coupled systems). It is shown that the
SU(3) spin-orbit-coupling gives rise to qualitatively different
phenomena and in particular we find that even a homogeneous SU(3)
field on a simple square lattice enables a topologically non-trivial
state to exist, while such SU(2) systems always have trivial
topology. In deriving this result, we first establish an 
equivalence between the Hofstadter model with a $1/N$ Abelian flux per
plaquette and a homogeneous SU$(N)$ non-Abelian model. The former is
known to have a topological spectrum for $N>2$, which is thus
inherited by the latter. It is explicitly verified by an exact
calculation for $N=3$, where we develop and use a new algebraic method
to calculate topological indices in the SU(3) case. Finally, we
consider a strip geometry and establish the existence of three gapless
edge states -- the hallmark feature of such an SU(3) topological insulator.
\end{abstract}
\maketitle

Following the theoretical prediction \cite{kane05,bernevig06} and
experimental observation \cite{konig07} of
the quantum spin hall effect, topological states of matter have
received a recent surge of attention.  The classification of topological
states of matter lies outside of the Landau symmetry breaking
paradigm, and is instead determined by topological quantum numbers \cite{wen04}.
The existence of nonzero topological numbers often has important physical
consequences for finite systems, including the existence of edge states
\cite{halperin82}.
Strong spin-orbit coupling is central to the
experimental realization of the quantum spin hall effect.
Spin-orbit coupling has a long history in solid-state systems and 
can play a number of important roles \cite{winkler03}.
Recently, interest in spin-orbit coupling has come to the fore in the
seemingly disparate area of ultracold atoms with the advent of
synthetic gauge fields \cite{lin09a}.  Such gauge fields have been
employed to mimic magnetic fields \cite{lin09,aidelsburger11,
struck12, garcia12} as well as Rashba-Dresselhaus spin-orbit coupling
in both bosons \cite{lin11} and fermions \cite{cheuk12}.  This
progress opens doors not only to quantum simulation of spin-orbit
coupled solid-state systems \cite{mazza12,bloch12}, but also to the
realization of a much larger class of  structures that can be
engineered in the ultracold laboratory but do not exist in the solid
state (see, for instance, \cite{dalibard11, cooper11, anderson12}).

In this Letter, motivated by the recent advances in ultracold atoms,
we introduce the notion of SU$(N)$ spin-orbit coupling.  In
particular, we consider a system of atoms in a square optical lattice
under the presence of spin-orbit coupling corresponding to a spatially
homogeneous gauge field, as such gauge fields are experimentally
simpler to realize \cite{dalibard11}.  We show that for $N=2$ (the
case most relevant in the solid state) all such Hamiltonians are
topologically trivial.  On the other hand for $N=3$ (as can be
realized with ultracold atoms with internal spin degrees of freedom
\cite{stamper12} but is less relevant for solid state systems),
by direct construction we show that such systems with nontrivial
topological numbers exist.  This topological property
results in the physically interesting situation of gapless edge modes,
while the bulk spectrum remains gapped.  Such modes can be
experimentally probed through \emph{in situ} imaging
\cite{Stanescu09}, time-of-flight spectroscopy \cite{Zhao11}, or Bragg
Spectroscopy \cite{Goldman12}.  The experimental realization of the
fairly simple resulting three-component Hamiltonian would pave the way
to the realization of topological states of matter in the ultracold
laboratory.

The conventional spin-orbit coupling in solid state systems manifests
itself as a Zeeman magnetic field that depends on the electron's
momentum \cite{winkler03}.  Hence a
typical spin-orbit term in a continuum model of a solid is ${\bf
b}({\bf p}) \cdot \hat{\bm \sigma}$, where the form of the
momentum-dependent internal field, ${\bf b}({\bf p}) $, is dictated by
symmetries of the crystal structure and $\hat{\bm \sigma}$ is a vector
of Pauli matrices, which mathematically are generators of the SU(2)
group that act on the electron's SU(2) spin. In contrast to solids,
synthetic spin-orbit structures in ultracold atoms are built from the
ground up and are not constrained by fundamental
symmetries. Furthermore, since the ``spin'' itself is synthetic, there
is no requirement that it be associated with a representation of the
SU(2) group. Hence, a much larger space of SU$(N)$ spin-orbit
couplings become available for multicomponent atoms, $\sum_{i}
b^i({\bf p}) \hat{X}_i$, where $\hat{X}_i$ are in principle any of the
$(N^2 -1)$ Hermitian generators of SU$(N)$ (e.g., the Pauli matrices for $N=2$,
the Gell-Mann matrices for $N=3$, etc.).

The Bloch Hamiltonian for a square lattice with nearest neighbor
hopping under the presence of a homogeneous  SU$(N)$ gauge field
is given by
\begin{equation}
\label{Eq:BlochH}
\hat\Hc(\bk)=-2t \left[ \cos(\kx - \hat{A}_x) + \cos(k_y - \hat{A}_y) \right],
\end{equation}
where $t$ is the hopping and the gauge fields $\hat{A}_{x,y}$ are
constant $N\times N$ Hermitian matrices.  To make the connection with
spin-orbit coupling clear, the cosines can be expanded and the
Bloch Hamiltonian can be rewritten as $\hat\Hc(\bk) = a(\bk) + \sum_i
b^i(\bk) \hat{X}_i$.  To construct an ``SU(3) topological insulator'' we relate the
model~(\ref{Eq:BlochH}) to the Hofstadter model \cite{hofstadter76},
familiar from quantum Hall physics, 
which describes particles hopping on a square lattice under a uniform
magnetic field (but non-uniform gauge field).  
To extend the Hofstadter model to SU$(N)$
systems we consider  $N$ decoupled replicas, each having the same flux
per plaquette.  Our starting point is thus the Hamiltonian
\begin{equation}
\label{Eq:Hof}
H_{\rm \tiny HM}=-t \sum_{i} \left(\Psi_{i}^\dagger \Psi_{i+\xh}+ \Psi_{i}^\dagger
e^{-i2\pi \alpha (x_i + \hat{S}_z)}   \Psi_{i+\yh} + {\rm
  H.c.} \right).
\end{equation}
In this equation $\Psi_{i}=(\psi_{i1}, \psi_{i2},\ldots,\psi_{iN})^T$
are SU$(N)$ spinor operators, $\hat{S}_z={\rm diag}(s,s-1,\ldots,-s)$ where
$2s+1=N$, $\xh$ and $\yh$ are the two square lattice vectors where the
lattice constant is set to unity,  $x_i=\xh\cdot {\bf r}_i$ where
${\bf r}_i$ is the position of the $i$\textsuperscript{th} lattice
site, and $\alpha$ gives the magnitude of the flux.  Since, $\hat{S}_z$ is
diagonal in this representation, the model trivially decouples into
$N$ independent copies of the Hofstadter model, each having
$2\pi\alpha$ flux per plaquette.  We restrict the flux to be related
to the number of spin components as $\alpha=1/N$.

We will first illustrate the mapping for the case of two-component
spins and later describe how to generalize.  For this case we apply the
gauge transformation
$
\Psi_i \rightarrow
e^{-i\frac{\pi}{2} \hat{\sigma}_x x_i}
\Psi_i
$ 
where $\hat{\sigma}_x$ is a Pauli matrix.  This transformation rotates the
spinors about the $x$-axis by a position-dependent angle.  As can be
seen after some straightforward algebra, this transformation removes
the spatial dependence of the second term in Eq.~(\ref{Eq:Hof}) at the
cost of introducing a non-Abelian $x$-component into the gauge field.  In
particular, after the gauge transformation Eq.~(\ref{Eq:Hof}) becomes
\begin{align}
H &=-t \sum_{i} \left(\Psi_{i}^\dagger e^{-i\hat{A}_x} \Psi_{i+\xh}+ \Psi_{i}^\dagger
e^{-i \hat{A}_y}  \Psi_{i+\yh} + {\rm
  H.c.} \right) \notag
 \\
&= \sum_{\bk} \Psi_\bk^\dagger \hat{\Hc}(\bk) \Psi_\bk,
\end{align}
where in the second line we have taken the Fourier transform.  For
this case, the
non-Abelian gauge fields of Eq.~(\ref{Eq:BlochH}) can be expressed as
Pauli matrices as $(\hat{A}_x,\hat{A}_y)=\frac{\pi}{2}(\hat{\sigma}_x,\hat{\sigma}_z)$.  The
Bloch Hamiltonian can also be expanded and rewritten as $\hat{\Hc}(\bk)=-2t[
\sin(\kx) \hat{\sigma}_x + \sin(k_y)\hat{\sigma}_z]$ which is a lattice version of
Rashba spin-orbit coupling.

We now generalize this mapping to  any integer $N$.  As before, we
perform a gauge transformation $\Psi_i \rightarrow \hat{\Uc}^{x_i} \Psi_i $.
The unitary matrix $\hat{\Uc}$ is defined to have $\hat{\Uc}_{1,N}=\hat{\Uc}_{n+1,n}=-i$
for $1\le n \le N-1$ with zeroes elsewhere.  One can verify for this
matrix that $\hat{\Uc}^\dagger e^{-i2\pi\alpha \hat{S}_z}\hat{\Uc}= e^{-i2\pi\alpha
(\hat{S}_z-1)}$.  Therefore this gauge transformation will completely remove
the position dependence of the second term in Eq.~(\ref{Eq:Hof}), and
the transformed Hamiltonian will correspond to particles on a square
lattice under a homogeneous gauge field.  

The SU(3) case of this general mapping will be considered in detail
below.  For this case, the non-Abelian gauge fields arrived at through
the mapping which enters Eq.~(\ref{Eq:BlochH}) can be expressed in
terms of Gell-Mann matrices \cite{georgi99} as
\begin{equation}
\label{Eq:model}
\hat{A}_x = \frac{2\pi}{3\sqrt{3}} ( \hat{\lambda}_2 - \hat{\lambda}_5 +
\hat{\lambda}_7) \mbox{ and }
\hat{A}_y  = \frac{\pi}{3} (\hat{\lambda}_3 + \sqrt{3} \hat{\lambda}_8). 
\end{equation}
By expanding the cosines, the Bloch Hamiltonian can also be written
as $\hat{\Hc}(\bk) = \bb(\bk) \cdot \hat{\blambda}$ where 
$
\allowbreak
\bb(\bk)=-t \left( \cos(\kx),  \right.
\sin(\kx),\frac{\sqrt{3}}{2}\sin(\ky) -\frac{3}{2}\cos(\ky) ,
\cos(\kx),-\sin(\kx),\cos(\kx),\sin(\kx),
\frac{\sqrt{3}}{2}\cos(\ky) +\frac{3}{2}\sin(\ky) \left. \right)
$
is an eight-component vector and $\hat{\blambda}$ is a vector composed of the
eight Gell-Mann $3\times3$ matrices.  

\emph{Geometrical Method for Berry Curvature and Chern Number
Computation}.  We now describe a geometrical method of computing the
Berry curvature and Chern numbers for general SU$(3)$ systems.  We first
write down expressions which are valid for any $N$.  The Berry
curvature $\Omega_\nc(\bk)$ \cite{berry84} is defined in terms of the
normalized eigenstates $\chi_{\bk \nc}$ of the Bloch Hamiltonian as
\begin{equation}
\label{Eq:Berry}
\Omega_\nc(\bk) = i \left( \partial_{\kx}
  \chi_{\bk \nc}^\dagger \partial_{k_y} \chi_{\bk \nc}-\partial_{k_y}
  \chi_{\bk \nc}^\dagger \partial_{\kx} \chi_{\bk \nc}
\right),
\end{equation}
where $\nc$ labels the eigenstate (or band).  
The Chern number  for a particular band is defined as
\cite{thouless82}
\begin{equation}
\label{Eq:thouless}
\nu_\nc = \frac{1}{2\pi} \int_{\rm BZ} d^2k \; \Omega_\nc(\bk),
\end{equation}
where the integral is performed over the first Brillouin zone (BZ).
The Berry curvature can also be expressed in terms of eigenstate
projection operators $\hat{P}_{\bk \nc} = \chi_{\bk \nc} \otimes
\chi_{\bk \nc}^\dagger$, where $\otimes$ denotes the outer product,
through the useful relation \cite{avron83,moore07}
\begin{equation}
\label{Eq:Avron}
\Omega_\nc(\bk) d\kx \wedge d\ky = i
{\rm Tr} ( \hat{P}_{\bk \nc} \wedge d\hat{P}_{\bk \nc} \wedge d\hat{P}_{\bk \nc}),
\end{equation}
where $d\kx \wedge d\ky = -d\ky \wedge d\kx$.

Before generalizing we first describe a well-known geometrical
expression for the Berry curvature for SU$(2)$ systems (see, e.g.,
\cite{hasan10}).  This will be used to demonstrate that SU(2) Bloch
Hamiltonians of the form Eq.~(\ref{Eq:BlochH}) are in general
topologically trivial.  The Bloch Hamiltonian for SU(2) systems can be
expressed in terms of Pauli matrices as $\hat{\Hc}(\bk) = a(\bk) + \bb(\bk)
\cdot \hat{\bsigma}$.  The projection operators corresponding to the two
eigenstates can be written in terms of $\bb(\bk)$ as
$
\hat{P}_{\bk \pm} = \frac{1}{2} \left[ 1 \pm  \bb(\bk)\cdot \hat{\bsigma} /|\bb(\bk)|  \right].
$
Inserting this into Eq.~(\ref{Eq:Avron}) then gives
\begin{equation}
\label{Eq:su2}
\Omega_\pm(\bk)= \mp \frac{1}{2|\bb(\bk)|^3} \bb(\bk) \cdot \left[
  \partial_\kx \bb(\bk) \times \partial_\ky \bb(\bk) \right].
\end{equation}
Thus, the Berry curvature can be expressed directly in terms of the
Bloch Hamiltonian, rendering the intermediate steps of computing its
eigenstates and evaluating Eq.~(\ref{Eq:Berry}) unnecessary.  
For SU(2) systems, one can write arbitrary gauge fields of $\hat{\Hc}(\bk)$
as linear combinations of Pauli matrices as $\hat{A}_{x,y}=u_{x,y}+{\bf
v}_{x,y}\cdot \hat{\bsigma}$.  After expanding the exponents to obtain
$\bb(\bk)$, it is a straightforward exercise to verify that
$\partial_\kx \bb(\bk) \times \partial_\ky \bb(\bk) \propto {\bf v}_x
\times {\bf v}_y$.  Then through Eq.~(\ref{Eq:su2}) one sees that the
Berry curvature vanishes identically, rendering SU(2) systems
described by Eq.~(\ref{Eq:BlochH}) topologically trivial.

We now move on to develop a central technical result of our work, namely the
generalization of Eq.~(\ref{Eq:su2}) to SU(3) systems.  We will utilize
the elegant formalism presented in \cite{khanna97} which describes an
efficient way to represent pure-state density matrices (or projection
operators) for three-state systems.  For SU(3) systems, a general
Bloch Hamiltonian can be expressed in terms of the eight Gell-Mann
matrices as
\begin{equation}
\label{Eq:Bloch3}
\hat{\Hc}(\bk)=a(\bk)+\bb(\bk) \cdot \hat{\blambda},
\end{equation}
where $a(\bk)$ is a scalar and $\bb(\bk)$ is an eight-dimensional real
vector.  The product of two Gell-Mann matrices can 
be written as
$
\hat{\lambda}_a \hat{\lambda}_b = 
\frac{2}{3} \delta_{ab} + 
d_{abc}\hat{\lambda}_{c} +i f_{abc} \hat{\lambda}_c
$
where $d_{abc}$ and $f_{abc}$ are the symmetric and antisymmetric
structure constants of SU(3) \cite{georgi99}.  These structure
constants define three bilinear operations for the eight-component vectors.
In particular, one has the dot product ${\bf u} \cdot {\bf v}= u_a
v_a$, the cross product $(\bu \times \bv)_a = f_{abc} u_b v_c$, and
the so-called star product \cite{khanna97} $(\bu * \bv)_a =
\sqrt{3} d_{abc} u_b v_c$ for two arbitrary vectors ${\bf u}$ and
${\bf v}$ where repeated indices are summed over.  One can also write
eigenstate projection operators in terms of the Gell-Mann matrices as
\begin{equation}
\hat{P}_{\bk \nc}= \chi_{\bk \nc} \otimes \chi_{\bk \nc}^\dagger= \frac{1}{3} ( 1 + \sqrt{3}
\bn_{\bk \nc} \cdot \hat{\blambda}).
\end{equation}
where ${\rm Tr}\, \hat{P}_{\bk \nc}=1$.  The condition that $\left(\hat{P}_{\bk
\nc}\right)^2=\hat{P}_{\bk \nc}$, leads to two constraints on the
vector $\bn_{\bk \nc}$ which are $\bn_{\bk \nc} \cdot \bn_{\bk \nc}
=1$ and $\bn_{\bk \nc} * \bn_{\bk \nc} =\bn_{\bk \nc} $
\cite{khanna97}.  Due to the star-product constraint, $\bn_{\bk
\nc}$ lies in a restricted region of $S^7$.  This can be compared
to the SU(2) system where the vector analogous to $\bn_{\bk \nc} $ can lie
anywhere in $S^2$.

Now we will express $\bn_{\bk \nc}$ in terms of $\bb(\bk)$
appearing in the Bloch Hamiltonian Eq.~(\ref{Eq:Bloch3}).  For
projection operators corresponding to eigenstates we have $[\hat{P}_{\bk
\nc}, \hat{\Hc}(\bk)]=0$ so that $\bb(\bk) \times \bn_{\bk \nc}=0$.
One can verify that this equation, along with the above constraints,
is satisfied by
$
 \bn_{\bk \nc}= \xi_{\bk \nc} \left[ \gamma_{\bk \nc}  \bb(\bk) +
 \bb(\bk) * \bb(\bk) \right]
$
with coefficients 
\begin{align}
\label{Eq:coeffs}
\gamma_{\bk \nc} &= 2 |\bb(\bk) | \cos\left(
 \theta_{\bk}+\frac{2\pi}{3} \nc \right);   \\
\xi_{\bk \nc} &=\frac{1}{|\bb(\bk) |^2\left[4
  \cos^2(\theta_{\bk}+\frac{2\pi}{3} \nc)-1 \right]},
\notag
\end{align}
where
$\theta_{\bk} = \frac{1}{3} \arccos\left[ \frac{\bb(\bk) \cdot
    \bb(\bk) *\bb(\bk) }{|\bb(\bk)|^3 }\right]$ and $\nc$ runs from
one to three.
The resulting expression for $\hat{P}_{\bk \nc}$ can be inserted into
Eq.~(\ref{Eq:Avron})
to obtain the Berry curvature.  One finds
\begin{widetext}
\begin{align}
\label{Eq:curve3}
\Omega_\nc(\bk)=- \frac{4\xi^3}{3^{3/2}} \left[\gamma^2 \partial_\kx \bb
  \times \partial_\ky \bb  + \gamma  \partial_\kx \bb
  \times \partial_\ky (\bb * \bb)
+  \gamma \partial_\kx (\bb*\bb) \times \partial_\ky \bb+ \partial_\kx (\bb*\bb) \times \partial_\ky (\bb*\bb)
\right]\cdot (\gamma \bb +  \bb*\bb),
\end{align} 
\end{widetext}
where we have suppressed the $\bk, \nc$ arguments on the right-hand
side.  Notice that due to orthogonality relations,
the derivatives do not act on the coefficients.  While
Eq.~(\ref{Eq:curve3}) is complicated in appearance, it is
straightforward to compute with a given $\bb(\bk)$.  This equation
provides an explicit expression for the Berry curvature in terms of
quantities from the Bloch Hamiltonian and thus should be viewed as a
generalization of Eq.~(\ref{Eq:su2}) to SU(3) systems.

\emph{Analysis of SU(3) model}. 
 Having established the above formalism, we now move on to analyze the
specific SU(3) model arrived at above, given by
Eqns.~(\ref{Eq:BlochH}) and (\ref{Eq:model}).  The resulting
$\bb(\bk)$ can be directly inserted into Eqns.~(\ref{Eq:coeffs})
and (\ref{Eq:curve3}) to find the Berry curvature for this system.
One finds
\begin{equation}
\Omega_\nc(\bk)=\frac{2\cos(4\theta_\bk+\frac{2\pi}{3}\nc)-3}{\sqrt{3}\left[1+2\cos(2\theta_\bk
-\frac{2\pi}{3}\nc)\right]^3 },
\end{equation}
where $\theta_\bk=\frac{1}{3} \arccos\left[  \frac{-1}{\sqrt{8}}\left(\cos(3\kx)+\cos(3\ky)\right)\right]$.
In addition, using the expression $E_{\bk \nc} = {\rm Tr}\{ \hat{P}_{\bk
  \nc} \hat{\Hc}(\bk) \}$, the bulk eigenenergies are found to be
\begin{equation}
E_{\bk \nc} = 2\sqrt{2} t
\frac{\cos(3\theta_\bk)+2\cos(\theta_\bk+\frac{2\pi}{3}\nc)}
{1+2\cos(2\theta_\bk+ \frac{2\pi}{3}\nc)}.
\end{equation}
These bands are gapped and ordered such that $E_{\bk 1} < E_{\bk 2} <
E_{\bk 3}$.  With the above expressions for the curvature, the Chern
numbers can be computed via Eq.~(\ref{Eq:thouless}) and are found to
be $(\nu_1,\nu_2,\nu_3) = (-3,6,-3)$.

\begin{figure}
\includegraphics[width=3.5in]{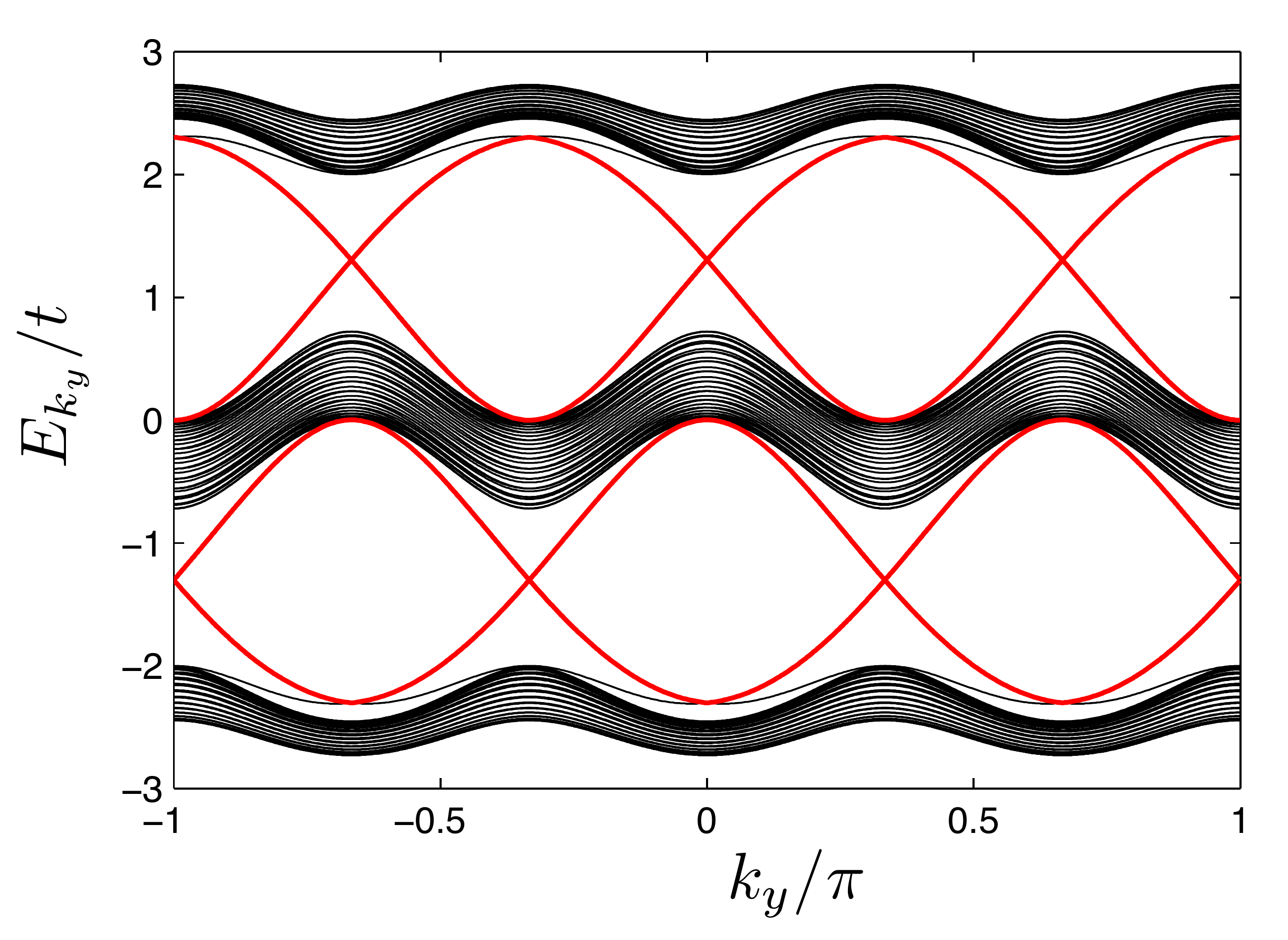}
\caption{The spectrum of SU(3) model in a strip geometry.
The bulk states correspond to black lines while the edge states
correspond to the thick red lines.  The presence of such topological
edge states is implied, through the bulk-boundary correspondence, 
by non-zero Chern numbers.}
\label{Fig:edge}
\end{figure}

Due to the bulk-boundary correspondence \cite{halperin82}, non-zero
Chern numbers imply the existence of edge states.  To elucidate the
behavior of these edge states, we investigate the SU(3) system in a
strip geometry.  We apply periodic boundary conditions in the
$y$-direction, and take a system of finite length in the
$x$-direction.  The system in this strip geometry is described by
\begin{align}
H_{\rm strip} & = -t \sum_{i}  \left[ \Psi_i^\dagger(\ky) 2\cos(k_y -
  \hat{A}_y) 
\Psi_i(\ky)    \right. \\
& \left. + \Psi_i^\dagger(\ky)e^{-i
    \hat{A}_x}\Psi_{i+1}(\ky)+\Psi_{i+1}^\dagger(\ky)e^{i \hat{A}_x}\Psi_{i}(\ky)
\right],\notag
\end{align}
where $i$ now is a one-dimensional finite sum.  The eigenstates of
$H_{\rm strip}$ are plotted in Fig.~\ref{Fig:edge}. The spectrum exhibits characteristic
topological edge states that connect the bands with different Chern numbers. 

In conclusion we make a few general remarks. First, we note that while
the SU(3) topological insulator constructed here relies on spin-orbit
coupling of a new type and while the calculation of Chern numbers
requires a new algebraic construction, its overall topological
characterization resides within the existing general classification
scheme \cite{Schnyder08,kitaev09} and corresponds there to a lattice
quantum Hall state labelled by an integer topological index.  However
in contrast to solid-state systems where the absence or presence of
time-reversal symmetry is an obvious physical constraint, for
synthetic spin-orbit systems the notion of time-reversal symmetry
does not have such a direct meaning, because the synthetic spins do
not behave like real spins under time reversal. Classification of
cold-atom Hamiltonians with respect to transformations of
time-reversal type can still be formulated but in a more formal way by
examining the existence of an anti-unitary symmetry of the Hamiltonian
which may or may not have a direct physical interpretation. From this
perspective, our Hamiltonian does not have such a symmetry.  One can
argue that in general such Chern topological insulators are much
easier to realize with cold atoms than $Z_2$ topological insulators,
because imposing an additional unphysical symmetry would require
fine-tuning the synthetic Hamiltonian, in contrast to the situation in
the solid state where in the absence of external magnetic fields and
magnetic impurities time-reversal invariance is automatically
preserved.  Finally, we briefly comment on experimental realization of
the SU(3) system.  There exists a considerable literature on the
realization of synthetic gauge fields in cold atom systems (for a
review, see \cite{dalibard11}). The gauge fields from
Eq.~(\ref{Eq:model}) can be realized with variations of the so-called
$N$-pod schemes \cite{Juzeliunas10, dalibard11}.  While the $N$-pod
schemes yield static gauge fields (as considered in this work) only,
there are proposed extensions to dynamical gauge fields
\cite{Banerjee12} whose study in the context of SU(3) systems have
important connections with particle physics and will be an interesting
avenue of future consideration.

{\em Acknowledgements --} This research was supported by JQI-PFC (RB),
ARO-MURI (GB), and US-ARO (VG).


%

\end{document}